\documentclass[sigconf, screen, nonacm]{acmart}

\AtBeginDocument{%
  }

\setcopyright{none} 
\author{Joshua Viszlai}
\authornote{Both authors contributed equally to this research.}
\authornote{Correspondance: viszlai@uchicago.edu}
\affiliation{%
  \institution{University of Chicago}
  \city{Chicago}
  \state{Illinois}
  \country{USA}
}
\author{Jason D. Chadwick}
\authornotemark[1]
\affiliation{%
  \institution{University of Chicago}
  \city{Chicago}
  \state{Illinois}
  \country{USA}
}
\author{Sarang Joshi}
\affiliation{%
  \institution{University of Chicago}
  \city{Chicago}
  \state{Illinois}
  \country{USA}
}
\author{Gokul Subramanian Ravi}
\affiliation{%
  \institution{University of Michigan}
  \city{Ann Arbor}
  \state{Michigan}
  \country{USA}
}

\author{Yanjing Li}
\affiliation{%
  \institution{University of Chicago}
  \city{Chicago}
  \state{Illinois}
  \country{USA}
}
\author{Frederic T. Chong}
\affiliation{%
  \institution{University of Chicago}
  \city{Chicago}
  \state{Illinois}
  \country{USA}
}

\newcommand{\circlenumber}[1]{\raisebox{.5pt}{\textcircled{\raisebox{-.9pt} {#1}}}}

\usepackage[normalem]{ulem}
\usepackage{amsmath}
\usepackage{amsthm}
\usepackage{xspace}
\usepackage{siunitx}
\usepackage[utf8x]{inputenc}

\begin{document}

\title{Predictive Window Decoding for Fault-Tolerant \\Quantum Programs
}

\begin{abstract}
Real-time decoding is a key ingredient in future fault-tolerant quantum systems, yet many decoders are too slow to run in real time. Prior work has shown that parallel window decoding schemes can scalably meet throughput requirements in the presence of increasing decoding times, given enough classical resources. However, windowed decoding schemes require that some decoding tasks be delayed until others have completed, which can be problematic during time-sensitive operations such as T gate teleportation, leading to suboptimal program runtimes. To alleviate this, we introduce a \emph{speculative} window decoding scheme. Taking inspiration from branch prediction in classical computer architecture our decoder utilizes a light-weight speculation step to predict data dependencies between adjacent decoding windows, allowing multiple layers of decoding tasks to be resolved simultaneously. Through a state-of-the-art compilation pipeline and a detailed simulator, we find that speculation reduces application runtimes by 40\% on average compared to prior parallel window decoders.
%SWIPER significantly reduces the temporal cost of using high-accuracy decoders, enabling better quantum error correction.
\end{abstract}


\maketitle

% Sections

\begin{figure}
    \centering
    \includegraphics[width=\linewidth]{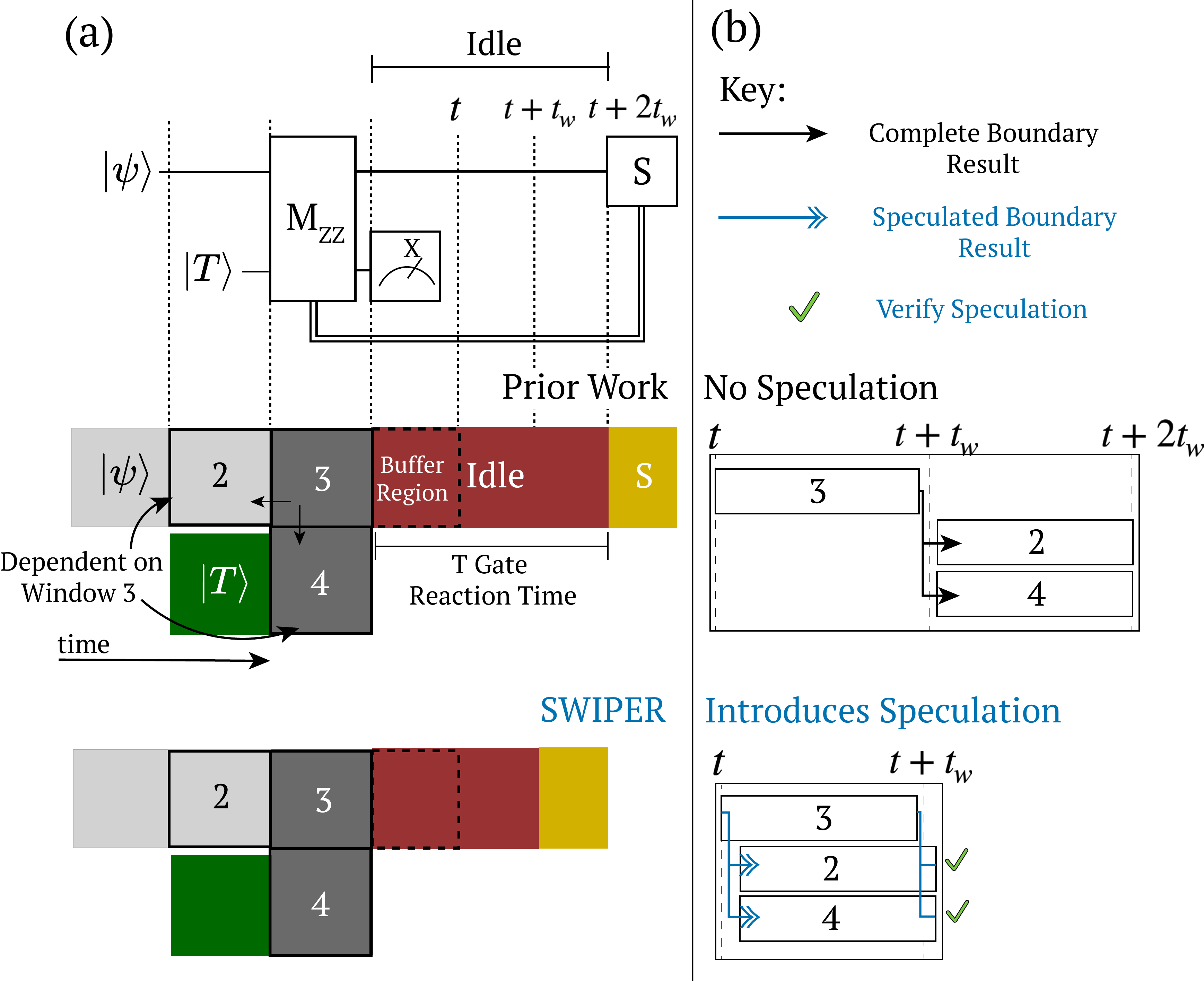}
    \caption{(a) Decoding a lattice surgery T gate teleportation (top left) using prior parallel window decoders (middle) and using SWIPER (bottom). 
    %The conditional S operation cannot be applied until the $\text{M}_\text{ZZ}$ operation is fully decoded. 
    (b) Resulting decoding pipelines. SWIPER improves decoding throughput with a lightweight \emph{speculation} step, allowing dependent windows to begin decoding earlier. 
    % \gokul{do you think it might be beneficial to first show the working of the speculative method in a simple single qubit example before moving to t injection? it still feels like readers might find it a little difficult to understand this figure even if they have understood Fig.1 well - for instance, the colored squares  in this figure are still hard to fully comprehend (though some additional text in the caption might help with that). More generally, readers might first br looking for the fig.1 equivalent of the proposed speculation method before they can understand its utility in a T injection scenario like this.} I changed the colors to make it easier to digest, but I think if we want to include a memory version we can do so in the background figure maybe?
    }
    % \gokul{dark vs light gray is unclear. also it might be worth defining what exactly is happening in 1,2,3 - maybe in the text if not in caption. Also one potential confusion for the reader might be that 3 executes before 1 but seems to be later in time compared to 1 (because it is to the right of 1) - so unless 1,2,3 are clearly defined its hard to reason about this}}
    \label{fig:speculation_intro}
\end{figure}

\section{Introduction}
Quantum computers are poised to deliver computational speedups for problems intractable classically. It is known that many of these problems, such as quantum chemistry and factoring, will require fault-tolerant systems to translate noisy physical qubits into resilient logical qubits via quantum error correcting (QEC) codes. A critical component in the realization of QEC is the \textit{decoder}, which must operate in real-time with the quantum device. Through classical algorithms acting on streams of parity check data, the decoder effectively tracks the state of error on logical qubits. A leading proposal for decoding a long-running quantum computation is to decode \textit{windows} of parity check data as they are generated. Adjacent windows must pass completed decoding data across window boundaries to create a global solution. In the first proposal from Dennis et al.~\cite{dennis2002topological}, known as sliding window decoding, data is passed forward in time, creating sequentially-dependent decoding problems.

While a decoding system is necessary for enabling QEC, its efficiency also has strong implications for the quantum computation. 
Terhal points out that if the rate of data production is higher than the rate of data decoding, then the computation will experience an \emph{exponential} slowdown~\cite{terhal2015quantum} in the sliding window setting due to an accumulating backlog of decoding tasks that all rely on their predecessors. In response, research has broadly improved decoder performance through a combination of algorithmic innovations~\cite{delfosse2021almost, higgott2023sparse, wu2023fusion} and tailored optimizations~\cite{ravi2023better, vittal2023astrea, alavisamani2024promatch}. When addressing the backlog problem, however, it is important to clarify decoder \textit{throughput} versus decoder \textit{latency}. Decoder throughput is the rate at which information is decoded, whereas decoder latency is the time taken to actively decode a single problem. Innovations and optimizations in latency are important and they will, in turn, increase throughput. However, as discussed both by Skoric et al.~\cite{skoric2023parallel} and Tan et al.~\cite{tan2023scalable} the decoding backlog is chiefly a problem of throughput. The \emph{parallel window decoding} strategies that they propose show how throughput requirements can be scalably met while remaining agnostic to the underlying decoding latency. In parallel window decoding, the direction of data movement between windows is modified, partitioning windows into two alternating layers. Windows in the same layer are independent and can therefore be decoded in parallel. This removes long dependency chains inherent to sliding window decoding, and as a result, given enough classical decoders to operate in parallel, the throughput of the decoding system can be arbitrarily high, effectively resolving the backlog problem. 
%\jason{make cost of running program on a QPU 40\% less by reducing runtime; cite IonQ or similar to show that price is very high per second}

Although prior parallel window decoding schemes successfully meet throughput requirements, their impact on the time from when a decoding window is generated to when it is decoded, known as the \emph{reaction time}, is sub-optimal. %\gokul{do we also want to define throughput and latency in the previous paragraph - it may not be immediately clear to the reader as to what the difference between decoder latency and reaction time is}. Good point. I just added this in!
Ideally, the decoding of each window should begin as soon as the window is generated from the device. However, as shown in the decoding pipelines of Figure~\ref{fig:speculation_intro}b, in prior implementations a window which depends on data from an adjacent window cannot begin until the adjacent window is complete, increasing the reaction time.
%\gokul{maybe this should say, 'as per prior imlementations...'? also, the previous paragraph mentions that sequential dependence is removed and this line mentions that there is some dependence - this might be confusing to the readers - perhaps the previous pragraph should add a little more clarity without getting into details?}. Good point. I tried to better clarify the background concept in the prev paragraph
If the time to decode a window is $t_w$, the reaction time becomes at least $2t_w$. 
%depending on the structure of its dependencies with other already-existing or not-yet-generated windows.
This is particularly detrimental for ``blocking" operations, such as non-Clifford gates (e.g. T, CCZ), which require the system to be fully decoded before the program can continue. An increase in reaction time in this context causes additional idle operations to be inserted until decoding is complete, slowing down the computation.

%\jason{reference classical architecture ideas more} 
In this work we introduce SWIPER to reduce decoding reaction time in window decoding schemes. SWIPER translates speculation strategies commonly used in classical computer architectures to the problem of window decoding, removing unnecessary idle time waiting for data dependencies as shown in Figure~\ref{fig:speculation_intro}. 

To perform impactful speculation SWIPER leverages key insights: \circlenumber{1} Since data dependencies between adjacent decoding windows only exist along the window boundaries, they only constitute a fraction of the total decoding problem. For example, at a code distance $d=21$, the boundary is only $\sim 1/21 = 4.8\%$ of a window's total syndrome data. \circlenumber{2} Error chains crossing the boundaries between windows will be short and sparsely distributed in an overwhelming majority of cases. In such cases, solving for data dependencies does not need a full-fledged decoding process. 
With these insights in mind
%,we introduce a fast, light-weight predictor to speculate the data dependencies between adjacent decoding windows. 
SWIPER's predictor solves the simpler problem of finding short-weight matchings across the boundary, which can be done much more efficiently than a full decoding of the window. This tentative result can then be forwarded to the adjacent window, allowing it to begin decoding. Importantly, this does not replace the full decoder, which is still run lazily 
%\gokul{can we say this is run 'lazily' or something similar} Good thought!
and verifies the speculation's correctness. 

We find SWIPER provides a $40\%$ reduction in fault-tolerant program runtime compared to prior work, a critical improvement given the high demand and time unit cost of quantum processors which will need to be run on the order of hours to days for fault-tolerant programs~\cite{gidney2021factor, beverland2022assessing}. Our results utilize our decoding simulator, SWIPER-SIM, that allows simulation of window decoders for fault-tolerant lattice surgery programs. 
% We find SWIPER, shown in Figure~\ref{fig:speculation_intro}, reduces the time quantum programs wait for blocking operations by $40\%$ compared to prior work. 

% We find our proposed 3-step predictor has an accuracy $>90\%$ for code distances up to 25 within \SI{1}{\us} on an FPGA. 
% %We also develop SWIPER-SIM, a roudn-level lattice surgery window decoding simulator.
% %to examine the impact of SWIPER on fault-tolerant benchmark programs. 
% In representative fault-tolerant program benchmarks, we find that SWIPER gives a consistent $\sim 40\%$ reduction in program runtime compared to prior parallel window decoders.

%\textcolor{Red}{Discuss more here?}

\subsubsection*{The main contributions of SWIPER are as follows:}

\begin{itemize}
    \item We introduce SWIPER, a new windowed decoder that leverages a lightweight, FPGA-compatible predictor to \textbf{improve decoder reaction time by up to $\mathbf{50\%}$} compared to prior parallel window decoders.
    \item We develop \textbf{SWIPER-SIM, a round-level lattice surgery decoding simulator}, enabling program-level simulations of parallel window decoding and allowing us to analyze how decoding reaction time impacts overall benchmark runtime.
    \item Using SWIPER-SIM, we discover variance in reaction time for prior parallel window decoders based on the alignment of T gates. We therefore introduce an \emph{aligned window schedule} to enforce proper alignment, \textbf{improving reaction time by up to 50\% in alignment-limited settings}.
    \item We study SWIPER under varying decoder latency and show that \textbf{for fixed runtime constraints, SWIPER relaxes decoder latency requirements consistently by over $\mathbf{2-5\times}$,} enabling the development of more powerful and accurate decoders.
    \item We evaluate SWIPER with realistic decoder parameters on an assortment of representative benchmarks and \textbf{demonstrate consistent program runtime reductions of $\mathbf{40\%}$ regardless of program size}.
    \item We analyze the added classical overhead of SWIPER and provide a heuristic to determine how many classical decoders to allocate for SWIPER. We find the benefits of SWIPER come at the cost of a consistent $31\%$ increase in the number of concurrent decoders compared to prior parallel window decoding.

    % \item We show precisely how decoding reaction time impacts overall program runtime in a windowed decoder, and we show how to minimize this reaction time by optimally assigning regions to windows.
    % \item \jason{(TODO) speculation of windows with SWIPER}
    % \item \jason{(TODO) (open source?) full lattice surgery + decoding simulator; first work to fully specify edge cases of windowing methods; fully general and can be applied to any quantum circuit, and runs efficiently enough to look at large(ish) circuits}
    % \item \jason{(TODO) use realistic distribution of decoding times, and show how the spread affects runtime (as opposed to a constant decoding time)}
    % \item \jason{(TODO) evaluate on quantum algorithms, with largest consisting of X operations; results: XX-XX (geomean XX) improvement in runtime across all benchmarks.}
\end{itemize}
\begin{figure*}
    \centering
    \includegraphics[width=0.9\textwidth]{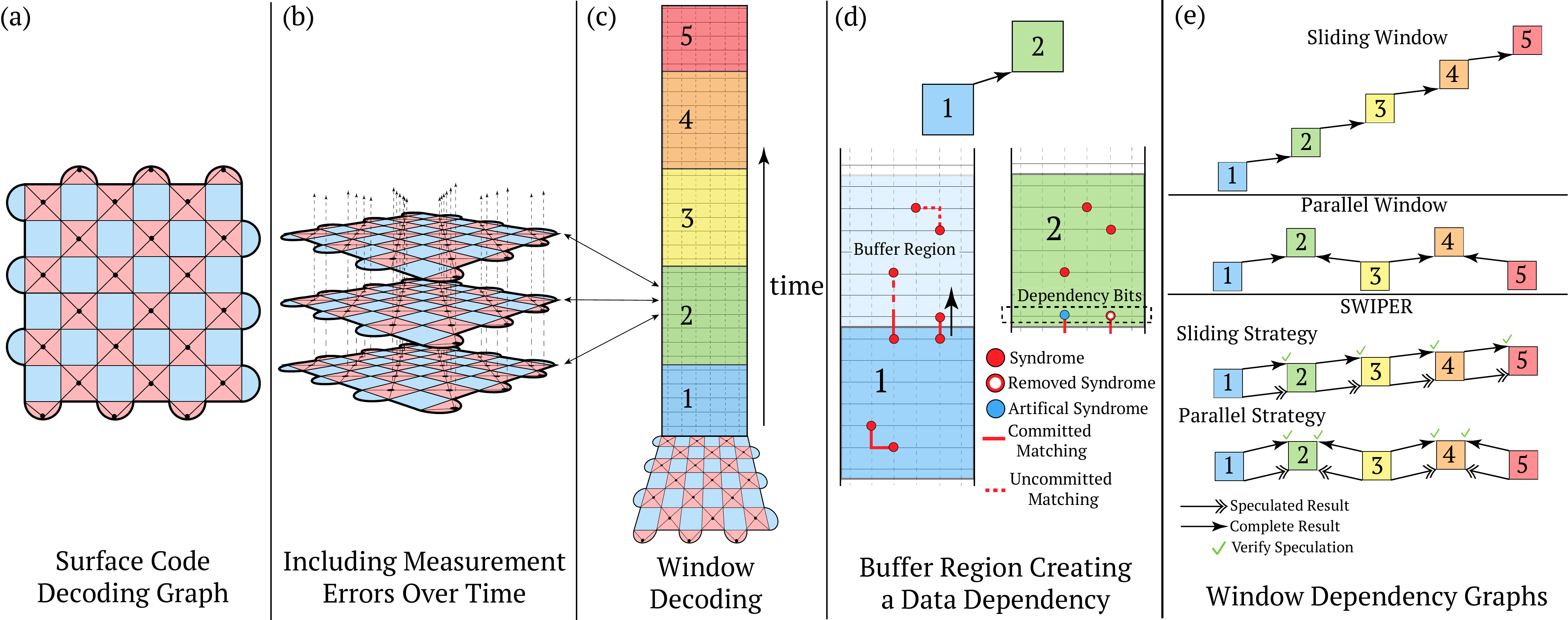}
    \caption{(a) A $\mathbf{d=7}$ surface code patch with Z(X) stabilizers indicated by red(blue) faces. Drawn is the decoding graph for Z stabilizers, with vertices of the graph indicating stabilizers and edges indicating error-prone data qubits. (b) Repeated rounds of measuring stabilizers. The decoding graph is augmented to include a temporal dimension with new edges indicating possible measurement errors. (c) A simplified view of decoding a surface code over time with one spatial dimension omitted. The decoding graph is split into windows labeled 1-5 which are each treated as a separate decoding problem. 
    (d) Decoding of window 1 and its buffer region. Matchings in the window are committed and create Pauli frame updates while matchings crossing into the buffer region create either artificial syndromes or removed syndromes at the boundary. Dependency bits on the boundary are passed to window 2 and therefore constitute a data dependency between windows 1 and 2 (denoted by an arrow).
    (e) Dependency graphs in a sliding window strategy and parallel window strategy. Arrows indicate the presence of a buffer region belonging to the window at the source of the arrow. 
    %\gokul{this might be hard but is it possible to show a quantification of time in this figure. i.e. to illustrate that swiper takes less time}  Noted! I'll definitely try and do this
    Also included are the same dependency graphs but when using SWIPER. Speculated dependencies (double arrows) allow SWIPER to start decoding windows sooner, and full decoding information is later used to verify the speculation (green check mark). }
    %\gokul{explicitly labeling the light blue portion as the buffer region might be useful } \jason{I think this part is fine}. Good thought, added label of buffer region! 
    \label{fig:background}
\end{figure*}

\section{Background}

For in-depth background we refer to~\cite{ding2020quantum, nielsen2001quantum, rieffel2011quantum} for quantum computing fundamentals and~\cite{fowler2012surface,gottesman1997stabilizer} for QEC and stabilizer codes. In this section, we give necessary background on the decoding problem and prior windowed decoding approaches.

\subsection{Surface Codes}
The surface code, shown in Figure~\ref{fig:background}a, is a leading QEC code that fits in a 2D grid with nearest-neighbor connections. As a CSS code, it has two sets of stabilizers ($Z,X$) for decoding bit-flip errors and phase-flip errors respectively, with each treated as a separate decoding problem. Operations on the surface code are typically formulated as lattice surgery~\cite{horsman2012surface}, in which adjacent surface code patches are \textit{merged} and \textit{split} to perform logical operations. These primitives enable universal computation~\cite{fowler2018low, litinski2019game} which has led to a growing set of resource estimates for large quantum programs~\cite{beverland2022assessing, gidney2021factor}.

\subsection{Quantum Error Correction Decoding}
We presume decoding takes place in the context of a long-running set of surface code patches, each generating syndrome data from their stabilizers. A decoding algorithm then operates on the syndrome data and aims to produce the most likely combination of physical errors that explains the observed syndromes. If all possible errors flip either a pair of stabilizers or a single stabilizer, as is the case with the surface code, we can use \textit{matching} decoders~\cite{fowler2012towards, delfosse2021almost, higgott2022pymatching}. In a matching decoder, we construct a decoding graph where nodes are stabilizers and edges are error mechanisms, such as data qubit errors and measurement errors. For the surface code, this graph is a 3D lattice, shown in Figure~\ref{fig:background}b. Decoding then consists of matching nodes with non-zero syndromes together to find a viable explanation for the observed syndromes. If the overall matching of syndromes is minimum-weight, it is a good approximation for the most likely error.

\subsection{Blocking Operations}
Decoding results can be tracked in software for some time using Pauli frames~\cite{knill2007quantum, riesebos2017pauli}, omitting the need to apply immediate corrective operations. However, non-Clifford gates in the program constitute \emph{blocking} operations which we cannot commute our Pauli frame past. Instead, a Clifford correction based on up-to-date decoding results must be applied physically in order to continue past the blocking operation~\cite{chamberland2018fault}. For surface code computation, it is common to use T as the only non-Clifford gate, applied via a T gate teleportation circuit. In this case, the result of the measurement operation in the teleportation circuit must be fully decoded before the S gate correction can be applied~\cite{divincenzo2007effective, terhal2015quantum}. In T-based computation, the delay between this measurement operation and the conditional S correction is based on the decoder's \emph{reaction time}, as shown in Figure \ref{fig:speculation_intro}.
The presence of blocking operations therefore necessitates \textit{real-time decoding}. In order to progress our quantum computation past blocking operations, a classical decoder must operate in conjunction with the quantum device. 

\textbf{\emph{Key Insight:} A delay in decoding a blocking operation causes a delay in the execution of a whole quantum program, so minimizing this reaction time is critical to ensure fast program runtimes}.

\subsection{Decoding Windows}\label{sec:back_window}
The latency of matching-based decoding algorithms grows polynomially with the decoding graph size, which places a practical limit on the size of a decoding problem. Instead of operating on the entire decoding history prior to each non-Clifford gate, we can instead operate on a pipeline of smaller \textit{windows} of decoding data, shown in Figure~\ref{fig:background}c. In the original proposal, termed the overlapping recovery method~\cite{dennis2002topological}, each window has a commit region and a buffer region, as shown in Figure~\ref{fig:background}d. In order to preserve the error-correcting performance of the underlying code, the buffer region should span $\sim d$ rounds, where $d$ is the code distance. The commit and buffer regions together constitute a decoding task that is sent to the ``inner'' decoder.
After a window is decoded, the matchings in the commit region are used to update the Pauli frame. 
Matchings that cross the boundary between the commit region and the buffer region can create "artificial" syndrome bits on the boundary that are passed to the next window. Similarly, any matchings from the commit region onto the boundary itself may remove a syndrome bit.
The set of syndrome bits on the boundary between the commit and buffer regions therefore contain information that needs to be passed to the next window. We will refer to these bits as the \textit{dependency bits}. 

In general, buffer regions are needed to pass information related to whether a potential error string could cross the boundary between the commit regions of adjacent windows. 
%For settings such as lattice surgery, it is also necessary to consider decoding windows across space as well as time due to the presence of long, merged surface codes during routing. 
A decoding window can also have multiple buffer regions if it is adjacent to multiple windows in space and time, which may occur during lattice surgery~\cite{lin2024spatially}. %\gokul{this para is hard to comprehend} I tried to rewrite to be more clear, lmk if it's ok.
At a boundary between two adjacent windows, a choice must be made for which window contains the buffer region (and so must be decoded first). Its result is then passed to the other window via the dependency bits. To capture this, we can define each boundary in a decoding window's commit region as a ``source" or ``sink". Prior work has also referred to these as ``rough" and ``smooth" boundaries, respectively. Source boundaries are followed by a buffer region and pass information to the adjacent window, while sink boundaries receive the passed information. 

\textbf{\emph{Key Insight:} The choice of boundary types in each decoding window enforces an ordering of decoding problems and therefore has an important impact on the overall decoding performance.}

\subsection{Sliding Window Approach}
The overlapping recovery method~\cite{dennis2002topological} is an example of \emph{sliding window decoding}. Examining a single surface code over time, each window always starts with a sink boundary and ends with a source boundary, meaning data is always passed sequentially. As shown in Figure~\ref{fig:background}e, these dependencies mean that each window cannot begin decoding until the previous window is fully decoded. The decoding throughput is therefore equivalent to the decoding latency, so in order to avoid a backlog, the latency must be lower than the time to generate a new window. 

We note that, to our knowledge, prior work has not examined the construction of spatially-sliding windows in the context of lattice surgery operations. For completeness, however, we will assume a sliding window decoder uses a similar, feed-forward approach for windows sliding along a spatial dimension.

\subsection{Parallel Window Approach}\label{sec:back_par}
In parallel window decoding~\cite{skoric2023parallel, tan2023scalable, lin2024spatially}, windows alternate between having all source boundaries and all sink boundaries. This minimizes the depth of the resulting dependency graph, shown in Figure~\ref{fig:background}e. Windows in the first layer have no data dependencies and can begin decoding immediately. Windows in the second layer are also independent from each other and can begin decoding after all of their dependencies are complete. For a long-running quantum program, this occurs in a pipelined manner, allowing many separate windows to be actively decoded in parallel. For spatially-parallel windowing of lattice surgery operations, an additional window type is needed that has a mix of source and sink boundaries \cite{lin2024spatially}.

\section{Motivation}

\subsection{Decoding Latency}
\begin{figure}
    \centering
    \includegraphics[width=\linewidth]{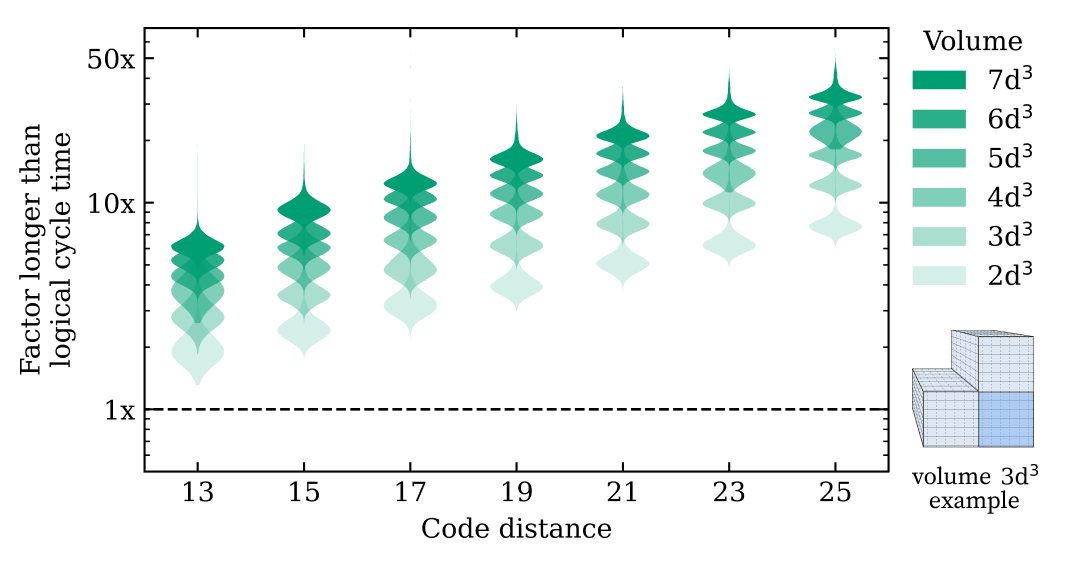}
    \caption{PyMatching decoding latency distributions for varying code distances and window sizes. Latencies are relative to a logical cycle consisting of $\mathbf{d}$ measurement rounds where each measurement round is assumed to take \SI{1}{\us}. \emph{Lower right:} example of a $\mathbf{3d^3}$ decoding volume in the style of Figure \ref{fig:background}c.}
    \label{fig:decoder_dists}
\end{figure}

The latency of an inner decoder is not necessarily fixed, and will generally vary with both the code distance and the specific decoding task (set of syndromes) at hand. Due to the complexity of decoding algorithms, in many cases the decoding latency will exceed the time to generate a decoding window. Regimes exist where this is avoided, particularly smaller code distances with hardware-based implementations~\cite{das2022lilliput, liyanage2023scalable, barber2023real}, however, in all of these the latency scales with the code distance $d$ with the highest distance achieved being $d=23$, still below distances expected for large-scale applications like factoring~\cite{gidney2021factor}. In Figure~\ref{fig:decoder_dists} we highlight this point by plotting decoding latencies using PyMatching, a state-of-the-art software decoder~\cite{higgott2022pymatching}, and Stim~\cite{gidney2021stim}, a detailed surface code simulator, with an assumed syndrome measurement round time of \SI{1}{\us} based on recent experiments~\cite{acharya2024quantum} and a physical error rate of $p=10^{-3}$. We can see latency scales not only with code distance but also the volume of the window, 
% \gokul{is window volume defined anywhere?}
which varies with the number of buffer regions as discussed in Section~\ref{sec:back_window}. Furthermore, this issue of latency is not unique to matching-based decoders. Higher accuracy decoders~\cite{bausch2023learning, shutty2024efficient} and decoders for qLDPC codes~\cite{panteleev2021degenerate} all struggle with decoding latency and therefore necessitate the use of an outer parallel window scheme to avoid a backlog. As such, we expect parallel window decoding to be a key ingredient in fault-tolerant quantum systems going forward. 

% \subsection{Window Decoding}\label{sec:window_motivate}
% \textcolor{red}{I will rewrite this and tie in Figure~\ref{fig:decoder_dists}.}
% Real-time decoding is a key requirement for future fault-tolerant quantum systems and will need to persist for exceedingly large code distances in order to enable expensive quantum applications such as factoring and chemistry. Many prior works on decoding have argued for prioritizing a latency below $1~\mu s$ as a requirement for real-time decoding~\cite{}, which is on par with recent experimental results for superconducting systems that generate a round of surface code data every $1.1~\mu s$~\cite{acharya2024quantum}. However this conflates decoding latency with throughput. Importantly, a decoding latency below $1~\mu s$ is \textbf{not} necessary to enable real-time decoding. Instead we assume a decoding latency larger than $1~\mu s$, which will be the case for large code distances and higher-accuracy decoders~\cite{}, where the use of recently introduced parallel window decoding is the leading approach for real-time decoding. In this framework, the underlying decoding algorithm operates as the "inner" decoder which is called for each decoding window allowing the window decoding scheme to be compatible with any inner matching decoder.

\subsection{Reaction Time}
We assume a quantum program running on a surface-code-based system will be decomposed into Clifford+T gates, where T gates constitute the only blocking operation. This is the leading approach for compiling to surface codes, with some synthesis strategies even omitting Clifford gates entirely~\cite{litinski2019game}. Since T gates are blocking operations, the reaction time of the decoder will determine how quickly T gates can be applied, and as a result, the overall program latency. 

In parallel window decoding, the reaction time of a decoding window is influenced by \circlenumber{1} the latency of the inner decoder run on each window and \circlenumber{2} delays caused by waiting for window dependencies. In this work, we address \circlenumber{2}, improving parallel window decoding in an inner-decoder-agnostic approach. With prior parallel windowing methods, \circlenumber{2} constrains the reaction time for windows with dependencies to be at least $2t_w$, where $t_w$ is the decoding latency. This is because a window with a dependency must wait until that dependency is \textit{completely} finished decoding before it can begin decoding. However, we argue this is not a fundamental constraint. In this work, we find that dependencies can be effectively predicted before the inner decoder is complete. SWIPER uses these speculations to reduce the impact of \circlenumber{2}. 

\textbf{\emph{Key Insight:} If the dependencies between windows can be resolved faster than the decoding latency, a window can begin decoding before its predecessors are complete.} 

% \subsection{Program Latency}
% We assume a quantum program running on a surface-code-based system will be decomposed into Clifford+T gates, where T gates constitute the only blocking operation. This is the leading approach for compiling to surface codes, with some synthesis strategies even omitting Clifford gates entirely~\cite{litinski2019game}. Since T gates are blocking operations, the reaction time of the decoder will determine how quickly T gates can be applied, and as a result, the overall program latency. 
% %To the best of our knowledge, there is no existing work analyzing the impact of decoder performance on fault-tolerant benchmark applications using a scalable window decoder. 
% To evaluate this, we introduce a simulation framework
% that allows us to evaluate SWIPER in the context of a complete fault-tolerant stack.

\section{SWIPER: Speculative Window Decoding}

In this section we introduce SWIPER, a window decoder that includes a light-weight prediction step to speculate data dependencies between adjacent decoding windows. We organize this section as follows: in Section~\ref{sec:t_inject} we describe how SWIPER changes the reaction time of T gates, in Section~\ref{sec:predictor} we outline a scalable predictor for surface code decoding, and in Section~\ref{sec:classical} we describe the classical decoder resources required by SWIPER. Then in Section~\ref{sec:eval} we describe the simulation methodology we use and present benchmark evaluations comparing SWIPER to prior, speculation-free window decoders.

\subsection{T Gate Teleportation}\label{sec:t_inject}
% T gates constitute a majority of operations in a fault-tolerant quantum program, with certain synthesis strategies even removing all Clifford operations thereby creating programs where \emph{all} operations require a T gate~\cite{litinski2019game, beverland2022assessing}. The performance of a quantum program therefore depends critically on the reaction time of a T gate.
Performing the T gate in the surface code requires the use of a T gate teleportation circuit, shown in Figure~\ref{fig:speculation_intro}. The S gate correction occurs $50\%$ of the time depending on the preceding measurement outcomes. Determining whether $S$ should be applied is blocking and requires the decoder to be up-to-date through the $ZZ$ merge operation (dark gray). As described in Figure~\ref{fig:background}e parallel window decoding requires that certain windows be dependent on future windows. This is seen in Figure~\ref{fig:speculation_intro} where windows 2 and 4 must wait to begin decoding until window 3 is completed at time $t+t_w$. In prior work this is unavoidable, however in SWIPER we speculate the dependencies, allowing windows 2 and 4 to begin decoding while window 3 is still decoding. After window 3 is complete, we then verify the speculations were correct. The resulting reaction time is reduced, allowing the program to continue past the blocking operation earlier than with prior work. Importantly, if we find that any speculations were incorrect upon completing window 3, the reaction time is still no worse than in prior work, because we can restart windows 2 and 4 at time $t+t_w$.

\textbf{\emph{Key Insight:} Window dependency speculation will never worsen reaction times compared to baseline methods and will generally improve them significantly.}

\subsection{Predictor Design}\label{sec:predictor}
% \newtheorem{theorem}{Theorem}[subsection]
% \newtheorem{lemma}[theorem]{Lemma}

% \gokul{probably not much space left for this but it would have been nice to have a figure that describes the predictor design}

In designing a predictor, we leverage the commonly-used fact that the majority of decoding problems will be simple with sparse, low-weight error chains~\cite{ravi2023better, vittal2023astrea, alavisamani2024promatch}. Since most error chains are low-weight, we can predict these dependencies by looking for small, simple patterns along the boundary. 
%This reduces the amount of syndrome data considered by $O(1/d)$. For example at $d=21$ we find examining between $4.8\%$--$9.5\%$ of the total syndrome data is sufficient for high-fidelity prediction.
% We aim to correctly predict this simple average case to achieve high prediction accuracy with minimal latency. This is in contrast to the full inner decoder, which must correctly handle worst-case decoding problems to ensure low logical error rates. 

%ignoring the majority of syndromes in the window and thus vastly reducing the computational cost compared to full decoding.

We develop our predictor design using an iterative approach. Here we describe how we iterate on a simple, 1-step predictor to create a 3-step predictor that handles the majority of common syndrome patterns that create data dependencies. We define the overall \emph{speculation accuracy} as the rate at which \emph{all} dependency bits along a boundary are predicted correctly; any single mistake constitutes an incorrect speculation.

\begin{figure}
    \centering
    \includegraphics[width=0.9\linewidth]{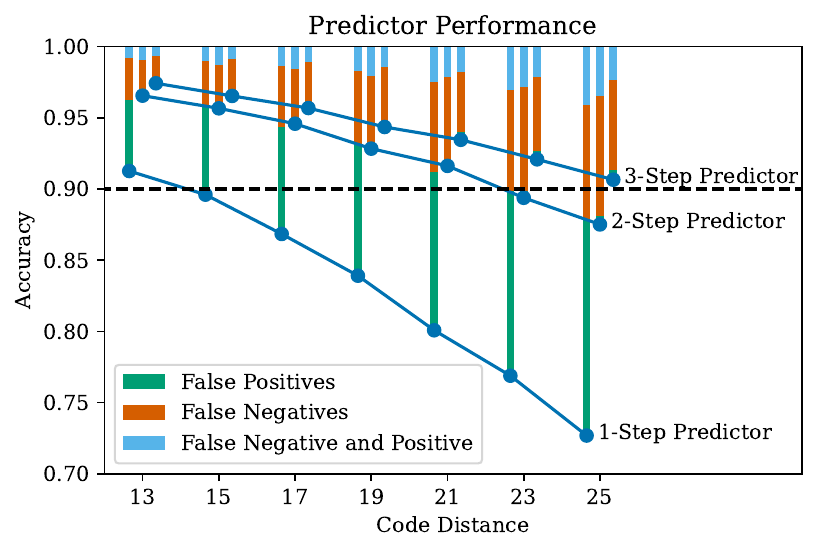}
    \caption{Accuracy of the three different predictors for increasing values of code distance. Bars indicate a breakdown of different failure cases. 
    %A false positive is when a matching crossing the boundary was predicted but did not actually exist and a false negative is when a matching that did cross the boundary was not predicted.
    }
    \label{fig:pred-accuracy}
\end{figure}

\begin{figure}
    \centering
    \includegraphics[width=0.6\linewidth]{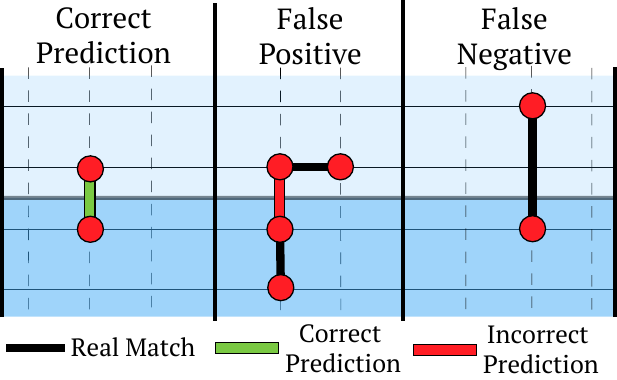}
    \caption{Common cases for the 1-step predictor. }
    \label{fig:1-step-fails}
\end{figure}

\subsubsection{1-Step Predictor}
In the 1-step predictor, we only look at single errors (edges in the decoding graph) that cross the boundary between a commit region and a buffer region. For each of these errors, the predictor checks their syndrome bits. If both bits are 1, it predicts that the error occurred and creates a dependency. We note that all errors can check their syndrome bits simultaneously, ensuring a low-latency, constant runtime prediction. Furthermore, for a distance $d$ surface code, the number of errors we need to check is only $O(d^2)$, which can be efficiently implemented in hardware logic.

In Figure~\ref{fig:pred-accuracy} we plot the accuracy of the 1-step predictor as well as a breakdown of misprediction cases. A false positive is when a matching crossing the boundary was predicted but did not actually exist and a false negative is when a matching that did cross the boundary was not predicted. We generate circuit-level surface code windows at a physical qubit error rate of $0.1\%$ using Stim~\cite{gidney2021stim} and we use PyMatching~\cite{higgott2023sparse} as the reference decoder to determine misprediction. Despite its simplicity, the 1-step predictor can still reach $> 70\%$ accuracy for all code distances we consider. 

The 1-step predictor can be further improved by studying common failure cases. In Figure~\ref{fig:1-step-fails} we describe the most common cases the 1-step predictor encounters.  While all failure cases are equally damaging and constitute a misprediction, from Figure~\ref{fig:pred-accuracy} we find that false positives are the most frequent mistakes made by the 1-step predictor.

\subsubsection{2-Step Predictor}
To address common false positives in the 1-step predictor, we propose an improved, 2-step predictor. The intuition from the 2-step predictor comes from the fact that after step 1 we may have clusters of errors that need to be pruned to create a minimal matching. We therefore design step 2 similarly to the peeling decoder introduced in~\cite{delfosse2020linear} where we prioritize matching syndrome bits with the fewest viable matches, akin to a leaf node.

We increase the set of errors considered to include all errors at most a distance of 2 from the boundary. In the first step, like the 1-step predictor, each error checks if both its syndrome bits are 1. If so, instead of declaring a match, they increment both syndrome bits by 1. In the second step, each error that believed itself a match is assigned to a bin based on the sum of its adjacent syndrome bits. Then in increasing bin order, each error checks if its syndrome bits are nonzero. If so, it declares itself a match and sets both syndrome bits to zero. Importantly, the number of bins is limited by the degree of vertices in the decoding graph which for the surface code is constant due to its constant, weight-4 parity checks. The runtime of this step is therefore independent of the code distance.
 % Furthermore, in the case where a cluster after step 1 is a minimum spanning tree for a set of syndrome bits, step 2 accomplishes the same operation as the peeling decoder.
%However since our goal is prediction rather than full decoding, failure cases are expected and acceptable so long as the prediction accuracy is sufficient. 
In Figure~\ref{fig:pred-accuracy} we find that the 2-step predictor resolves nearly all the false positive cases mispredicted by the 1-step predictor.

\begin{figure}
    \centering
    \includegraphics[width=\linewidth]{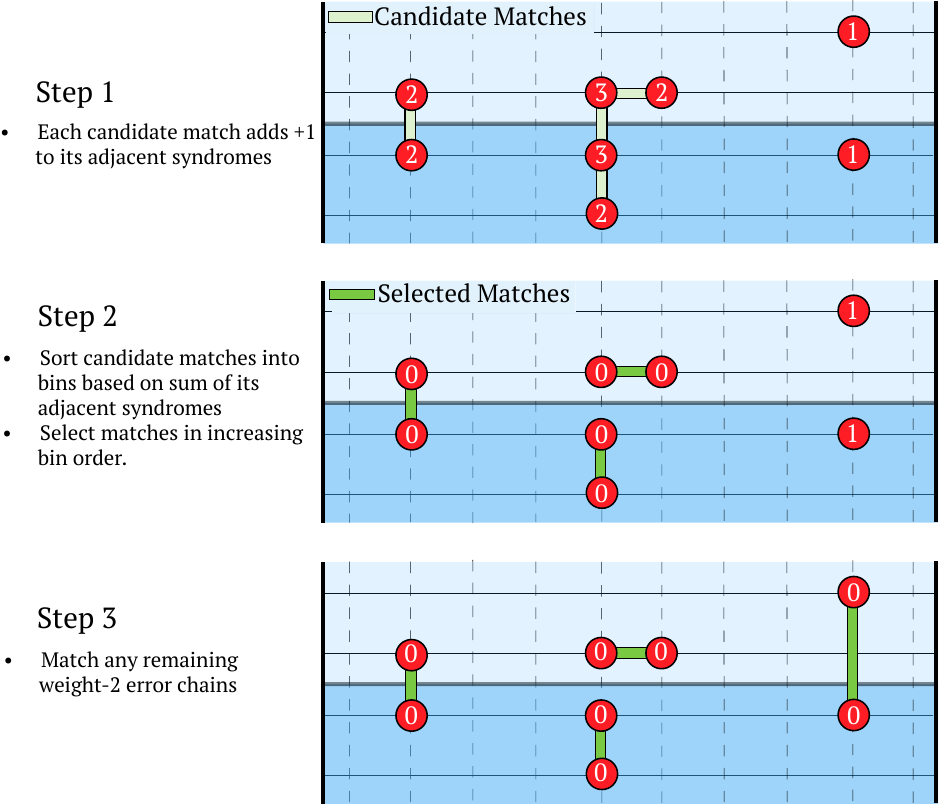}
    \caption{An overview of the final, 3-step predictor logic.}
    \label{fig:predictor-logic}
\end{figure}

\subsubsection{3-Step Predictor}
Finally, to address common false negative cases, we introduce a third step to the 2-step predictor. Offline, we can precompute the pairs of syndrome bits that can be matched by a weight-2 error chain crossing the boundary. Then, after the 2-step predictor determines its matches, we check each precomputed pair to see whether its syndrome bits are both 1 in the decoding graph, and if so, we declare a match. Similar to the 1-step predictor, these checks can all occur in parallel along the boundary ensuring a constant runtime with the number of such checks as $O(d^2)$. Since this occurs after the 2-step predictor, we expect all simple, weight 1 error chains to already be matched. We summarize the logic of this final, 3-step predictor in Figure~\ref{fig:predictor-logic}.

In Figure~\ref{fig:pred-accuracy} we can see the 3-step predictor reduces the amount of false negatives mispredicted by the 1 and 2-step predictors. As a result, we reach a prediction accuracy of $>90\%$ for all code distances we simulate. 

\textbf{\emph{Key Insight:} The 3-step predictor runs in time $O(1)$ since the number of unique bins in step 2 depends on the parity check weight, which is constant for the surface code.}

%\begin{table}[]
%    \centering
%    \begin{tabular}{c|r|r|r}
%         $d$ & Simulation time (ns) & \# of LUTs & \# of registers \\
%         \hline
%         13 & 60 \\
%         15 & 60 \\
%         17 & 60 \\
%         19 & 60 \\
%         21 & 60 \\
%         23 & 60 \\
%         25 & 60 \\
%         27 & 60 \\
%    \end{tabular}
%    \caption{FPGA predictor timing and cost}
%    \label{tab:fpga_eval}
%\end{table}

\subsubsection{Hardware Implementation}

To validate the behavior of the 3-step predictor, we implemented the algorithm on FPGA hardware. The FPGA environment was chosen as FPGAs have seen value as platforms for full QEC decoders~\cite{barber2023real, liyanage2023scalable}. We performed a behavioral simulation of the 3-step predictor on AMD's Vivado Design Suite~\cite{vivado} to verify the $O(1)$ runtime for $d=13$ through $d=27$. The time taken for the predictor implementation is always 60ns, demonstrating that it is constant with respect to distance. Since the 3-step predictor only considers $O(d^2)$ syndrome bits compared to the full decoding volume of $O(d^3)$, we also expect area costs to scale favorably compared to the full decoder. This separation increases as $d$ increases. For example, at the smallest window volume of $2d^3$, with $d=13$ the predictor examines $\sim 15\%$ of the total syndrome data whereas with $d=25$ the predictor only examines $\sim 8\%$ of the total syndrome data.

\begin{figure}
    \centering
    \includegraphics[width=0.8\linewidth]{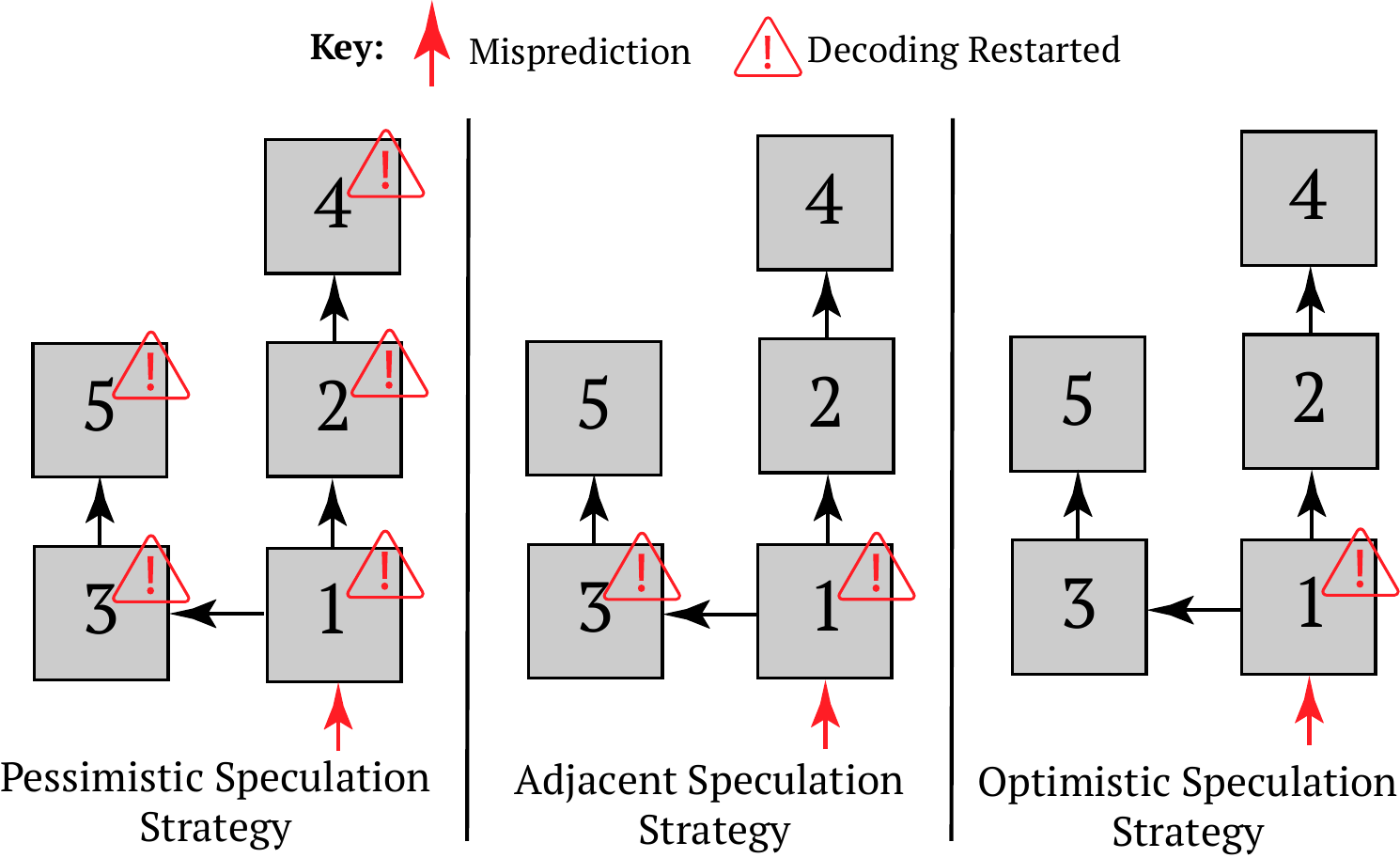}
    \caption{Three different speculation strategies for handling a misprediction which has poisoned window 1. ``Pessimistic'' restarts all descendents of a poisoned window. ``Adjacent'' restarts any window that used a prediction on an adjacent boundary to the misprediction. ``Optimistic'' restarts only the poisoned window itself. }
    \label{fig:poison_strategy}
\end{figure}

\subsection{Classical Resources}\label{sec:classical}
Parallel window decoding will need access to many classical decoders operating in parallel to keep up with throughput demands. Compared to prior, speculation-free schemes, SWIPER exhibits an increased demand on the number of concurrent decoders, as SWIPER collapses the dependency structure of windows. 
Given that parallel window decoding relaxes latency requirements, for this work we assume decoders exist out-of-fridge and are much less costly than the quantum device itself. As such, we believe an increase in classical compute to reduce the required amount of quantum compute is a beneficial trade. 
However, here we study how to minimize the required classical costs of SWIPER.
% In general we expect the decoding latency for a window to vary depending on the distribution and number of non-zero syndrome bits. \textcolor{red}{Needs 1-2 more sentences here}. In this work we presume SWIPER does not need to operate with limited classical decoders, however, studying window scheduling with limited decoder resources would be interesting for future work, particularly if considering costly, in-fridge decoders.
%Instead, we do study how to minimize the number of classical decoders required by SWIPER. 
Particularly, an incorrect speculation can cause wasted classical computation as it requires restarting some decoding tasks that were based on flawed dependency bits. SWIPER mitigates this cost through an \emph{optimistic speculation strategy}.

\begin{figure}
    \centering
    \includegraphics[width=0.7\linewidth]{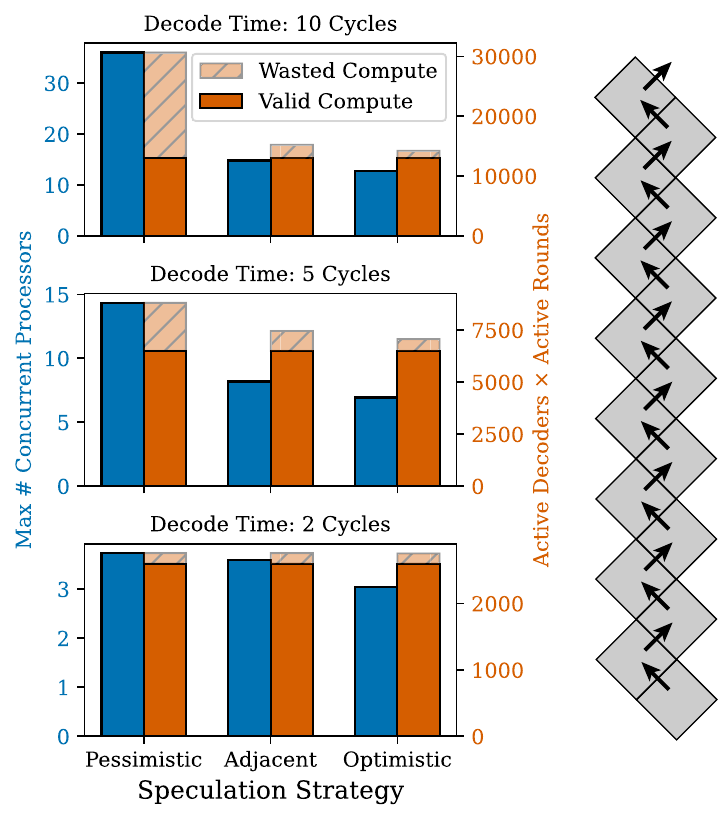}
    \caption{Estimating classical decoder costs of the three different proposed speculation strategies. All values are averaged over 10,000 shots. The specific case simulated is 100 windows in a ``zig-zag'' shape (right) using a sliding schedule with speculation.
    %We compare the maximum number of concurrent decoding processes (left axis) and the classical computation spacetime cost in terms of active decoders $\times$ active rounds (right axis). 
    In the breakdown, ``valid compute'' indicates compute that finished and was correct, whereas ``wasted compute'' indicates compute that was restarted early or found to be incorrect.}
    \label{fig:recovery_eval}
\end{figure}

\subsubsection{Handling Mispredictions}\label{sec:mispred}
A waiting decoding window begins decoding after its incoming dependencies have passed boundary speculations to it. Once each dependency is fully decoded, however, we must verify if the speculation was correct by comparing the dependency bits produced by the predictor and by the full decoder. In the event the predictor was incorrect, we have a \textit{misprediction}. A misprediction means that the relevant decoding window operated on incorrect syndrome data, leading to an invalid solution; the decoding task must be restarted using correct dependency bits. We call this window \textit{poisoned}. However, the effect of a misprediction does not necessarily stop here. Incorrect matching of syndromes along one boundary can lead to incorrect matching of other syndromes on other boundaries, so the effects of a misprediction may propagate further in the window dependency graph.
This raises an important question: which other decoding windows are unreliable in the event of a misprediction?

Given a dependency graph with a poisoned node, shown in Figure~\ref{fig:poison_strategy}, the \textit{pessimistic speculation strategy} (left) restarts all descendants that already began decoding. If the poisoned root significantly increases the chances of these descendants being poisoned, this strategy is effective, as it halts likely-incorrect decoding tasks earlier and avoids needless computation. On the other hand, if the descendants' chances of success are not significantly affected, this is a poor strategy, as it throws away valid, in-progress decoding tasks that are likely to be correct. As discussed in Section~\ref{sec:predictor}, we expect most decoding problems to be sparse with low-weight error chains. In this regime, we do not expect a change in syndrome bits on one boundary of a window to change a result on a different boundary. An \textit{optimistic speculation strategy} (Figure~\ref{fig:poison_strategy}, right) therefore only restarts the poisoned node itself. 

To evaluate this intuition, we use Stim and PyMatching to simulate the surface code at a physical qubit error rate of 0.1\% and estimate the conditional probability of a misprediction of a source boundary given a received misprediction on a separate sink boundary. We find no change in accuracy if the two boundaries are not adjacent (e.g. one temporal boundary to the next), and a $\approx 4\%$ decrease in accuracy if the two boundaries are adjacent (e.g. one spatial boundary and one temporal boundary). Given this asymmetry, we propose an intermediate strategy that restarts the poisoned node and any windows that used a prediction on an \emph{adjacent} boundary to the misprediction, shown in Figure~\ref{fig:poison_strategy} (center).

In Figure~\ref{fig:recovery_eval}, we show classical costs for decoding a sequence of 100 windows with speculation failures using the three speculation strategies detailed above. To study a regime where we expect the most variation between the three strategies, we use a sliding window schedule with a ``zig-zag'' dependency structure such that successive boundaries are always adjacent. We use a prediction accuracy of $90\%$ with a reduced accuracy of $86\%$ for boundaries adjacent to a misprediction. If the decode time is comparable to the time to generate a window (1 cycle) we see little variance, since the depth of the active dependency tree is small. However, at large decode times more similar to what we observed in Figure \ref{fig:decoder_dists}, we see that the pessimistic strategy unnecessarily restarts more windows, wasting classical compute. In all cases, we find the optimistic strategy performs the best, which we attribute to the high prediction accuracy and minimal downstream effects in the case of mispredictions. 

% \subsubsection{Inner Decoder Latency}
% As shown in Figure~\ref{fig:decoder_dists} the inner decoder latency will vary for different windows based on the syndrome complexity. As a result, windows may complete before their predecessors have completed and verified any speculations. Rather than wait for the predecessors to complete, SWIPER allows for an \textit{out-of-order commitment} of windows, freeing classical decoder resources as soon as possible. Importantly, this does not change the behavior of blocking operations. In the event of a blocking operation, all windows prior to the operation must still be completed, speculations verified, and Pauli frame updates committed before the program can continue. 

% If a committed window is later found to be poisoned, we must revert any Pauli frame updates and allocate a new decoder to restart its decoding. As a result, we must also store a record of each decoding problem until all speculations they used are verified. This record contains the device syndrome vector and Pauli frame updates.

\section{Methodology}\label{sec:eval}

To evaluate the impact of SWIPER on the overall program latency, we perform benchmark evaluations using state-of-the-art compilation techniques. 
To do so, we develop our own simulator, SWIPER-SIM, to evaluate window decoders given lattice surgery programs.
%To the best of our knowledge, SWIPER-SIM is the first software to simulate general lattice surgery programs in the context of window decoding.

\begin{figure}
    \centering
    \includegraphics[width=0.8\linewidth]{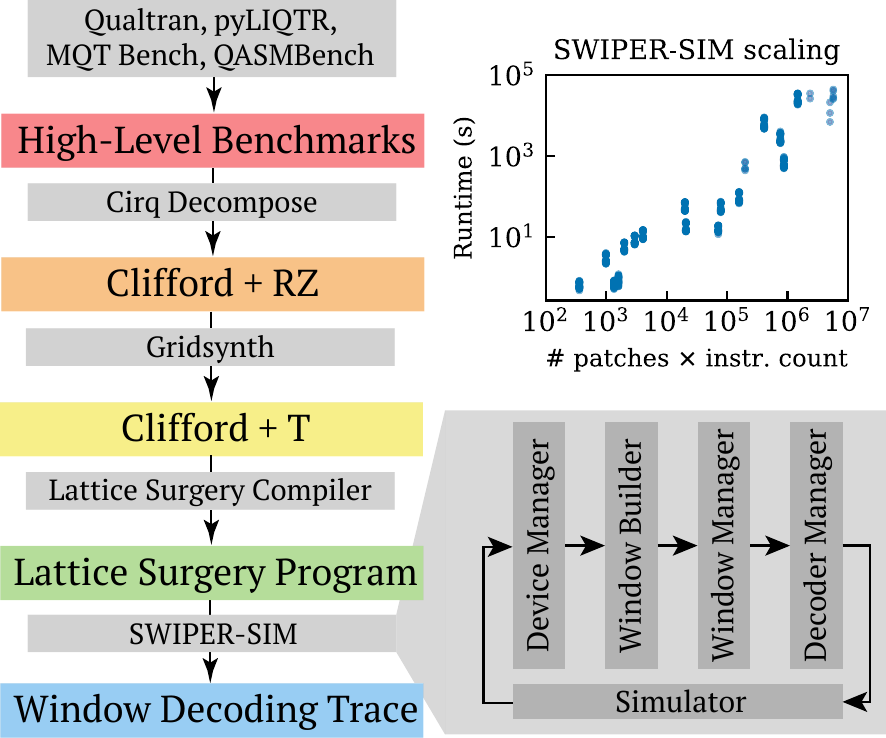}
    \caption{The pipeline used to evaluate high-level benchmark programs with different window decoders. \textit{Right:} details of SWIPER-SIM and its runtime versus program size on a single core of an Intel Xeon Gold 6248R processor.}
    \label{fig:simulation_overview}
\end{figure}

\begin{figure}
    \centering
    \includegraphics[width=\linewidth]{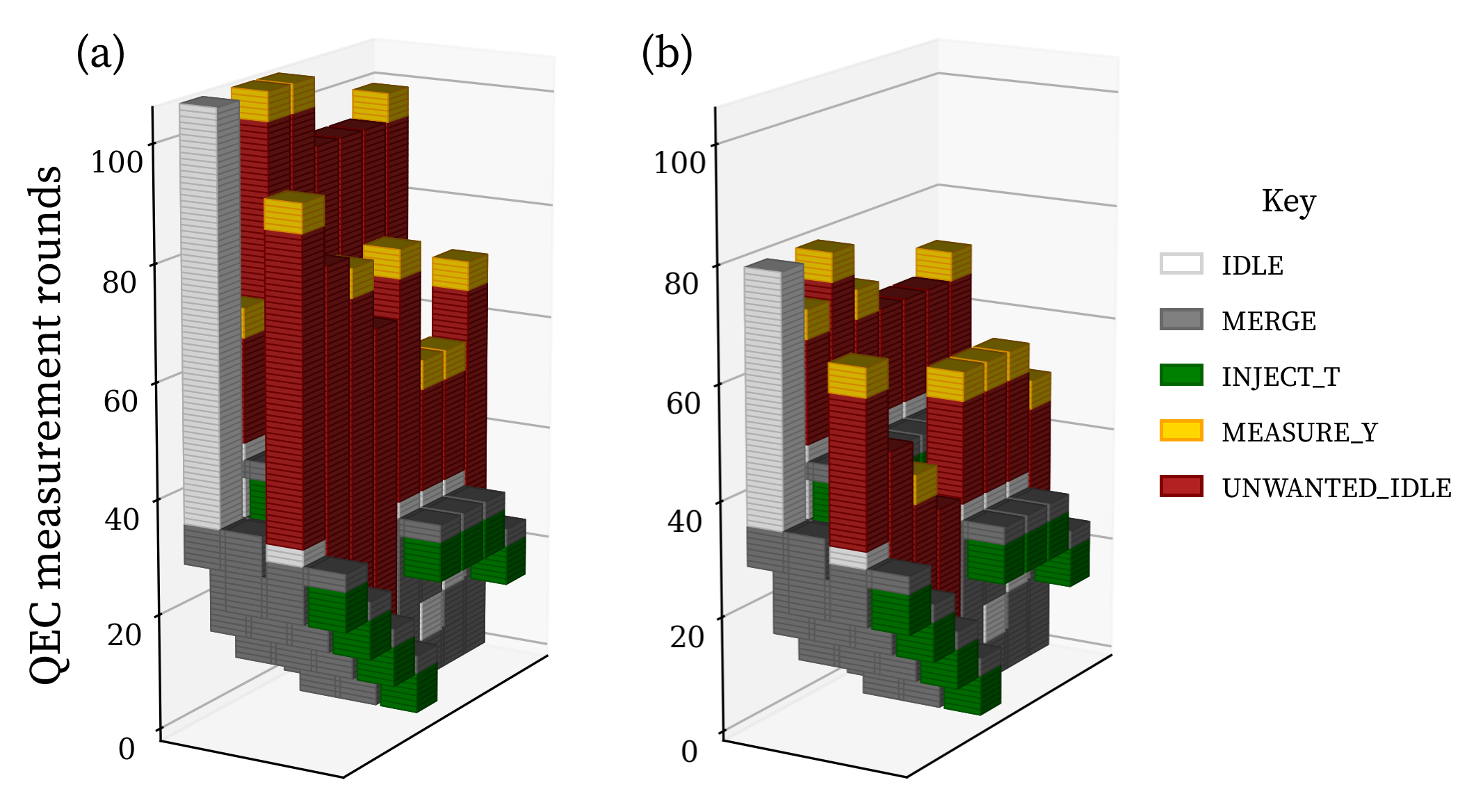}
    \caption{SWIPER-SIM program traces for a 15-to-1 magic state distillation in a distance-7 code using the construction from \cite{fowler2018low} (Fig. 17). Time advances vertically, and each horizontal slice represents a batch of syndrome data (colored by instruction type). Decoding time is fixed to be the same as the time to generate $\mathbf{2d}$ rounds, about twice the window generation rate. Device traces are shown for (a) baseline parallel window method ($\mathbf{15.1d}$ QEC rounds) and (b) SWIPER aligned window ($\mathbf{11.3d}$ QEC rounds, a 25\% improvement). 
    %SWIPER reduces the overall cost of the distillation subroutine by reducing dependency-induced latencies between successive decoding windows.
    }
    \label{fig:simulator-msd}
\end{figure}

\subsection{Simulation Software}\label{sec:sim_setup}
Figure~\ref{fig:simulation_overview} gives an overview of the procedure by which we compile and evaluate benchmark applications. We source relevant benchmarks for the fault-tolerant regime from recent repositories~\cite{harrigan2024expressing, pyliqtr, quetschlich2023mqtbench, li2023qasmbench} which can be compiled down to Clifford+RZ gates in OpenQASM~\cite{cross2022openqasm} using Cirq~\cite{cirq_dev}. We then use Gridsynth~\cite{ross2016optimal} to approximate RZ gates as a sequence of H, S, and T gates with a precision of $10^{-10}$. We feed the resulting Clifford+T circuit into the Lattice Surgery Compiler~\cite{leblond2024realistic} to create a mapped and routed lattice surgery program using the Edge-Disjoint Paths Compilation (EDPC) surface code layout~\cite{beverland2022surface}. 

SWIPER-SIM takes in a compiled lattice surgery program and performs a round-level simulation of syndrome generation, windowing, and decoding (the final step in Figure~\ref{fig:simulation_overview}). Based on the lattice surgery program, the DeviceManager generates a set of syndrome rounds every \SI{1}{\us} (based on recent experimental timing~\cite{acharya2024quantum}) for all currently-active surface code patches. These syndrome rounds are collected by the WindowBuilder and assembled into windows by the WindowManager, with source/sink boundaries decided based on a window decoding strategy (e.g. sliding, parallel). Completed windows are sent to the DecoderManager, which initiates speculation and decoding tasks when the relevant dependencies are satisfied, manages classical compute resources, and handles mispredictions. Speculation uses the optimistic strategy described in Section~\ref{sec:mispred} to handle mispredictions. For blocking instructions, the DeviceManager delays the conditional instruction until the DecoderManager signals that it has fully decoded (and verified) the instruction history up to and including the blocking operation.

By simulating at the granularity of rounds, windows do not necessarily need to align with instructions. As a result, idling while a blocking T gate is being decoded can complete as soon as possible, even if the reaction time is not a multiple of the window size. 
As an example, Figure~\ref{fig:simulator-msd} shows a spacetime program trace generated by SWIPER-SIM for a hand-specified lattice surgery program for 15-to-1 magic state distillation~\cite{bravyi2005universal, reichardt2004improved, fowler2018low}. We can compare prior work (Figure~\ref{fig:simulator-msd}a) with SWIPER (Figure~\ref{fig:simulator-msd}b) to see the reduction in reaction time for blocking operations when using speculation. Slices are colored by instruction type. Y basis measurement and S gates are modeled according to~\cite{gidney2024inplace}.
%Spacetime diagrams follow the same key as Figure~\ref{fig:speculation_intro}.
%, green denotes T injection, gray denotes lattice surgery merge and idles, yellow denotes the conditional S, and red denotes idles due to reaction time.

\subsection{Studying reaction times with SWIPER}\label{sec:align}

\begin{figure}
    \centering
    \includegraphics[width=\linewidth]{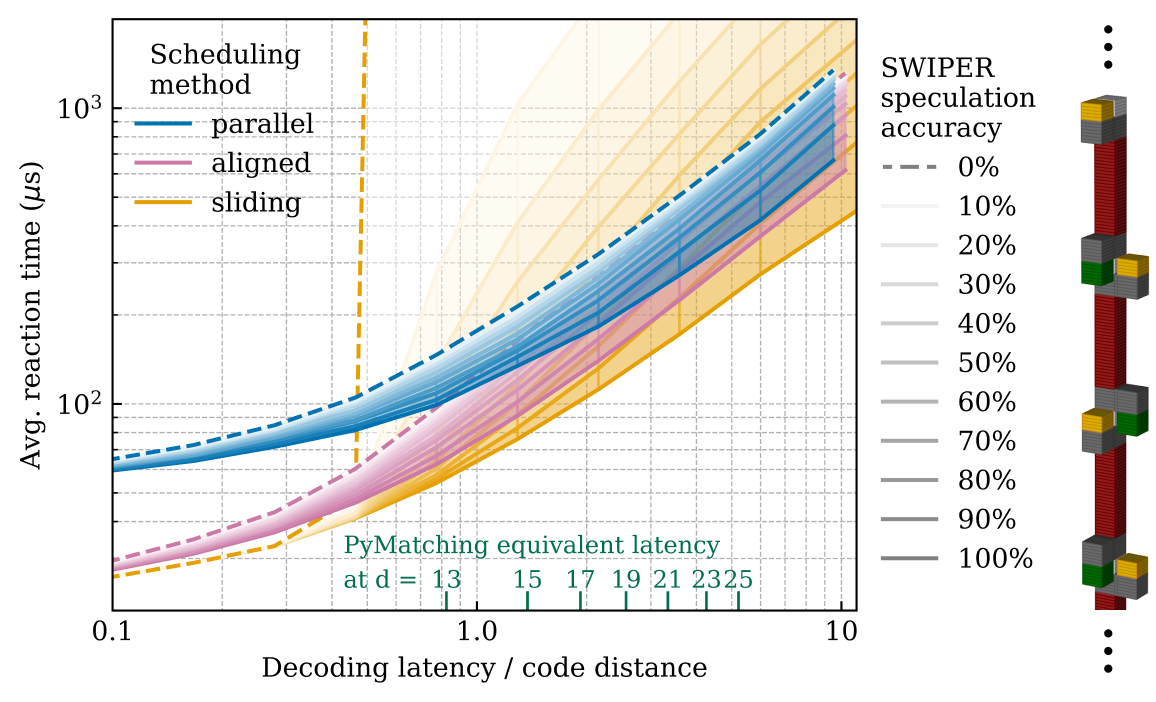}
    \caption{Sensitivity of reaction time to decoding latency and speculation accuracy. Speculation accuracy of $\mathbf{0\%}$ corresponds to baseline (no speculation) decoding. Decoder latency is $\mathbf{t_{\text{dec}}(v) = rv/d^2}$ rounds, where $\mathbf{v}$ is the volume of the decoding problem (in units of $\mathbf{d^3}$) and the relative latency factor $\mathbf{r}$ is varied along the x axis. Equivalent latency factors extracted from linear fits to PyMatching latency data (Figure \ref{fig:decoder_dists}) are shown along the x axis. \emph{Right:} simulator trace of repeated T gates on an idling logical qubit. A similar experiment with 1000 T gates was used to collect the data in this figure. 
    %\jason{look at how much ``headroom'' we enable in decoder latency for a fixed reaction time - cite Google slower decoder}
    }
    \label{fig:spec_accuracy}
\end{figure}

In Figure~\ref{fig:spec_accuracy} we plot the sensitivity of SWIPER to decoding latency and speculation accuracy. Decoding latency is proportional to window volume with relative factor $r$. For sliding windows of size $2d^3$ ($d^3$ commit, $d^3$ buffer), $r=0.5$ corresponds to a latency of $d$ rounds, matching the window generation rate; as expected, we see that the backlog problem for default sliding window decoding (yellow dashed line) therefore begins when $r>0.5$, where the reaction time begins to grow exponentially with each successive blocking operation. SWIPER mitigates this problem with speculation, but we find reaction time for sliding windows is particularly sensitive to speculation accuracy. This is due to the depth of the dependency tree, which is unbounded in sliding window decoding.

\subsubsection{T Gate Alignment}\label{sec:align}

\begin{figure}
    \centering
    \includegraphics[width=0.95\linewidth]{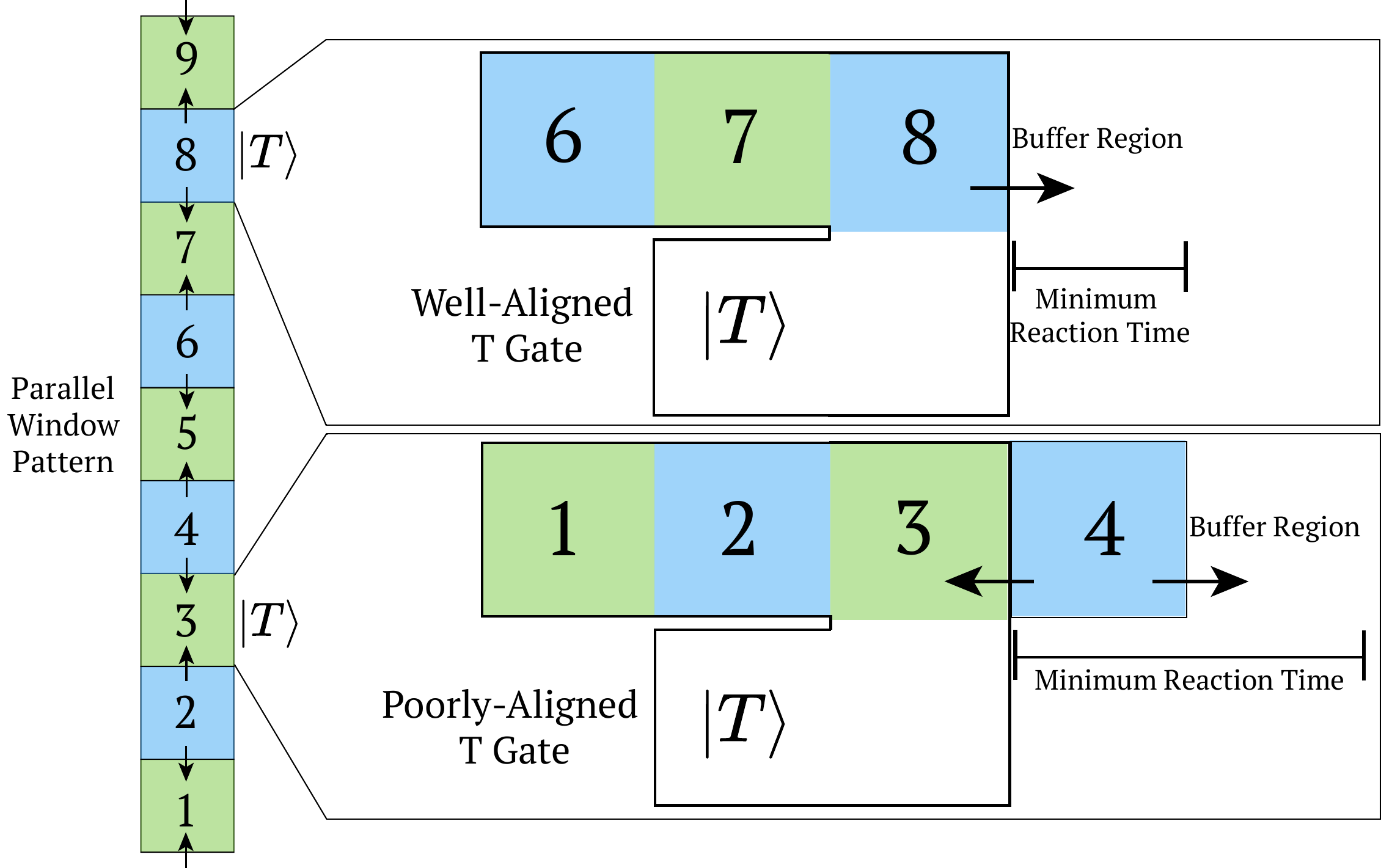}
    \caption{Comparing a well-aligned T gate (top) and a poorly-aligned T gate (bottom).}
    \label{fig:alignment_explain}
\end{figure}

We also see that at small decoding latencies the sliding window strategy outperforms the parallel window strategy by up to $50\%$. We can explain this in part by noticing that, in parallel window decoding, \emph{the type of window a T gate aligns with affects its reaction time}.
As shown in the bottom example of Figure~\ref{fig:alignment_explain}, if the merge operation in a T gate teleportation ends with a sink boundary, the reaction time can be further delayed by the time to generate the source boundary's window in the future. In this specific example, since window 3 depends on window 4, it must wait for window 4 to be generated 
%(and for the boundary to be speculated) 
before it can even begin decoding. However, if instead the merge ends in a source boundary, as shown with window 8 in the top example, the soonest window 8 can begin decoding is after its buffer region is generated, which is only $d$ rounds. 

More generally, we conclude that a blocking operation should always be ``aligned'' (have a future-facing source boundary) to ensure a minimal reaction time. Since a sliding window schedule always ends with source boundaries, it is always aligned. However the sliding window schedule is much more sensitive to speculation accuracy leaving it vulnerable to the backlog problem. To address this, we also introduce an \textit{aligned} window strategy, which is a parallel window strategy with forced alignment of blocking operations. 

\textbf{\emph{Key Insight:} Using SWIPER-SIM, we find that parallel window decoders suffer from dependency-induced delays, as already discussed, but also \textit{alignment}-induced delays.} 

\subsubsection{SWIPER's effect on reaction times}

In Figure \ref{fig:spec_accuracy}, we find the aligned strategy retains the misprediction resiliency of parallel window decoding while reducing reaction time. We see that for short decoding latency, the aligned strategy yields over $50\%$ shorter reaction times than the parallel strategy. As decoder latency increases, parallel and aligned strategies become increasingly similar; reaction time in this regime is dominated by waiting for dependencies to resolve rather than generating windows. For both aligned and parallel, we see that SWIPER can reduce reaction times by roughly $50\%$ when decoding latency is long; we attribute this to SWIPER effectively ``flattening'' the two-layer dependency structure of the parallel window strategy. A speculation accuracy to $90\%$ is sufficient to get most of the benefit of SWIPER.
Finally, we see that if the speculation is reliable enough, sliding window decoding performs best at all decoder latencies, which we attribute to the use of smaller window sizes (typically $2d^3$, compared to the $3d^3$-volume windows in parallel window decoding) leading to lower inner decoder latencies.

\subsubsection{Relaxing Inner Decoder Latency}
We can also analyze the data from Figure \ref{fig:spec_accuracy} from a different angle: in a setting with a fixed runtime budget, SWIPER allows for significantly longer inner decoder latencies. For fixed reaction time values, we compare high-accuracy SWIPER to the default parallel window scheme and find that speculated aligned and sliding schemes allow upwards of $5\times$ increased latency for reaction times near \SI{100}{\us} and all three (parallel, aligned, and sliding) speculated methods allow over $2\times$ increased latency in the limit of large reaction time (\SI{500}{\us}+). Again, we can understand this limit by remembering that SWIPER collapses the two-layer dependency structure of parallel windows.

This relaxation in decoder latency could be translated to an increase in program fidelity by using recurrent, transformer-based decoders~\cite{bausch2023learning}, tensor network decoders~\cite{chubb2021statistical, google2023suppressing}, or ensemble decoders~\cite{shutty2024efficient} which have demonstrated improvements to decoder accuracy at the cost of decoder latency.

\begin{table*}[]
    \centering
    \begin{tabular}{|l|l|r|r|l|l|}
        \hline
        Benchmark & Short Name & Space footprint & Compiled \# T's & Source & Domain \\
        \hline
        Toffoli & \texttt{toffoli} & 9 & 7 & \cite{nielsen2001quantum} & $\mu$Benchmark \\
        Magic State Distillation & \texttt{msd\_15to1} & 32 & 15 & \cite{fowler2018low} & Resources, $\mu$Benchmark \\
        RZ($\phi$), $\epsilon=10^{-10}$& \texttt{rz} & 4 & 100 & \cite{ross2016optimal} & $\mu$Benchmark \\
        \hline
        Quantum Read Only Memory & \texttt{qrom} & 153 & 48 & \cite{harrigan2024expressing} & Data I/O \\
        8-bit Adder & \texttt{adder\_8bit} & 111 & 112 & \cite{li2023qasmbench} & Factoring subroutine\\
        Carleman Encoding & \texttt{carleman} & 264 & 394 & \cite{pyliqtr} & ODE \\
        H$_{2}$ Qubitized Walk Operator & \texttt{H2\_molecule} & 92 & 3084 & \cite {pyliqtr} & Chemistry \\
        Fermi Hubbard $4\times4$ Lattice & \texttt{fermi\_hubbard} & 225 & 3898 & \cite{pyliqtr} & Chemistry \\
        Quantum Phase Estimation & \texttt{qpe} & 85 & 11378 & \cite{quetschlich2023mqtbench} & Common subroutine \\
        Quantum Fourier Transform & \texttt{qft} & 86 & 11484 & \cite{quetschlich2023mqtbench} & Factoring subroutine\\
        Grover's & \texttt{grover} & 34 & 13577 & \cite{quetschlich2023mqtbench} & Data Search \\
        \hline
    \end{tabular}
    \caption{Selected benchmark applications. Space footprint is the number of $d \times d$ surface code patches used.}
    \label{tab:benchmarks}
\end{table*}

\subsection{Benchmark Simulation}

To further evaluate the efficacy of SWIPER, we select a suite of fault-tolerant benchmark applications and use the methodology described in Section~\ref{sec:sim_setup} to simulate the program runtime when using different window decoding strategies. To evaluate at scales expected for large, fault-tolerant applications we consider a physical error rate of $p=10^{-3}$ and code distance $d=21$, which corresponds to logical error rates below $10^{-12}$. This is often referred to as the ``teraquop regime'' where a trillion logical operations can be executed. We again assume that each QEC syndrome round takes \SI{1}{\us} and we sample decoding latencies from the distributions shown in Figure~\ref{fig:decoder_dists}, rounding up if the window volume is not an integer multiple of $d^3$. Speculation is assumed to take \SI{1}{\us} with an accuracy of $90\%$ based on our results in Section~\ref{sec:predictor}.

\begin{figure*}
    \centering
    \includegraphics[width=0.9\linewidth]{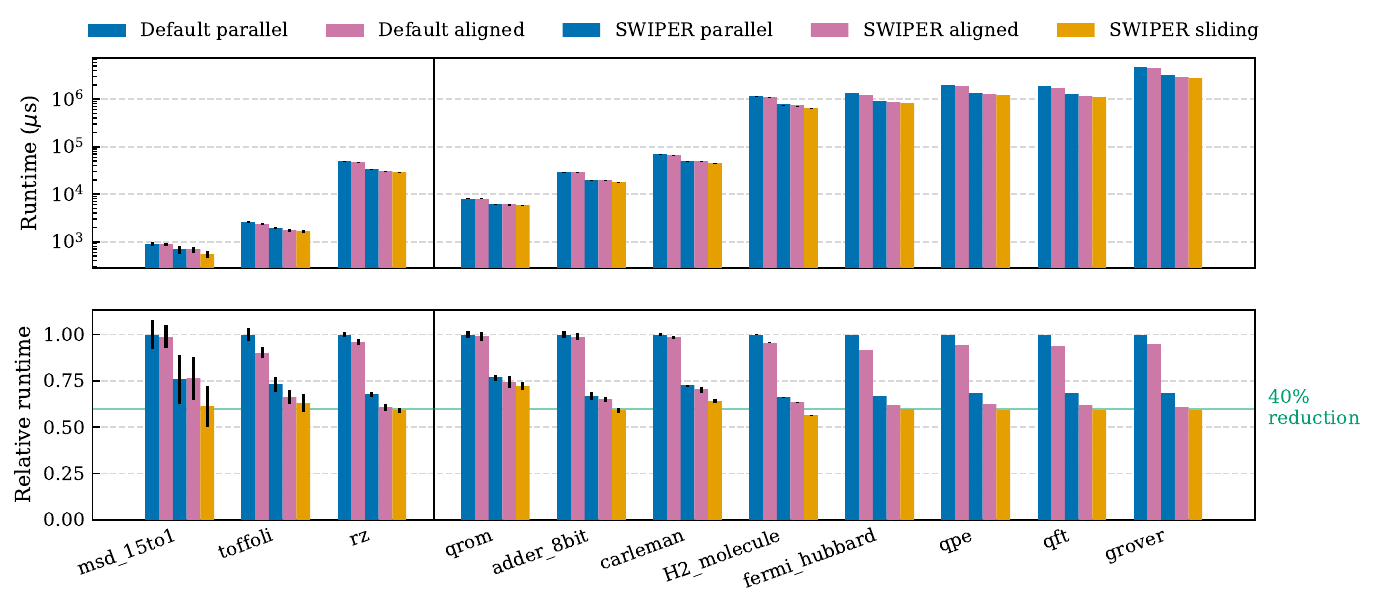}
    \caption{\emph{Top:} Program runtime for selected benchmarks under different window scheduling methods, with and without SWIPER. Classical processor resources are unlimited. Vertical black lines indicate standard deviation over 10 randomized trials for smaller benchmarks. 
    %\emph{Top right:} Relative differences in runtime when using auto-limited processor count (Section \ref{sec:classical_limit}) for benchmarks with runtime longer than $10^4$ \SI{}{\us}. 
    \emph{Bottom:} Relative runtimes normalized to default parallel window runtime.}
    \label{fig:benchmarks-combined}
\end{figure*}

\subsubsection{Selected Benchmarks}
Table~\ref{tab:benchmarks} summarizes the lattice surgery program benchmarks we simulate. We identify three ``microbenchmarks'' ($\mu$Benchmarks) which are small, consistently-used primitive operations in the fault-tolerant domain whose performance can be extrapolated to estimate that of programs beyond what we include in Table~\ref{tab:benchmarks}.

For the other benchmarks, we include exact quantum phase estimation on 5 qubits due to its presence as a common subroutine in many quantum algorithms. We include Quantum Read Only Memory (QROM) with 15 data bits and 15 select bits to represent data I/O in many quantum algorithms. Grover's algorithm on 5 qubits is chosen to represent data search applications. We include block encoding for Carleman linearization with 4 truncation steps to represent quantum algorithms for nonlinear differential equations.
We include the Quantum Fourier Transform (QFT) on 10 qubits and an 8-bit adder to represent subroutines in factoring applications. Finally, simulating the Fermi Hubbard model on a $4\times4$ lattice and performing qubitized ground-state energy estimation of $H_2$~\cite{babbush2018encoding} are included to represent chemistry applications.

% Since SWIPER impacts the reaction time of T gates, we find the number of compiled T gates is the most important program-level feature. 

\subsubsection{Results}

Figure \ref{fig:benchmarks-combined} shows the program runtimes for the selected benchmarks relative to the baseline parallel window method. Due to the uncertainty in runtime for smaller benchmarks (discussed in the following subsection), we report aggregate results for benchmarks with more than 1000 T gates: without SWIPER, using the aligned scheduling method achieves $4.3\%$ to $8.5\%$ (geomean $5.8\%$) reduction in runtime compared to parallel window. With SWIPER, all three scheduling methods (parallel, aligned, and sliding) improve significantly. SWIPER-parallel achieves $31.8\%$ to $33.9\%$ (geomean $32.7\%$) reduction in runtimes, SWIPER-aligned achieves $36.9\%$ to $39.2\%$ (geomean $38.1\%$) reduction in runtimes, and SWIPER-sliding achieves 40.4\% to 43.6\% (geomean 41.4\%) reduction in runtimes compared to baseline parallel window. We observe that the QROM and Carleman encoding benchmarks exhibit slightly less performance improvement compared to the other benchmarks; we conclude that this is because these two benchmarks have a lower fraction of T instructions ($\sim$40\% compared to $\sim$70-80\% for the other benchmarks), so the benefit of reducing reaction time is slightly less impactful.

\subsubsection{Runtime uncertainty and extrapolation}\label{sec:uncertain}

\begin{figure}
    \centering
    \includegraphics[width=0.9\linewidth]{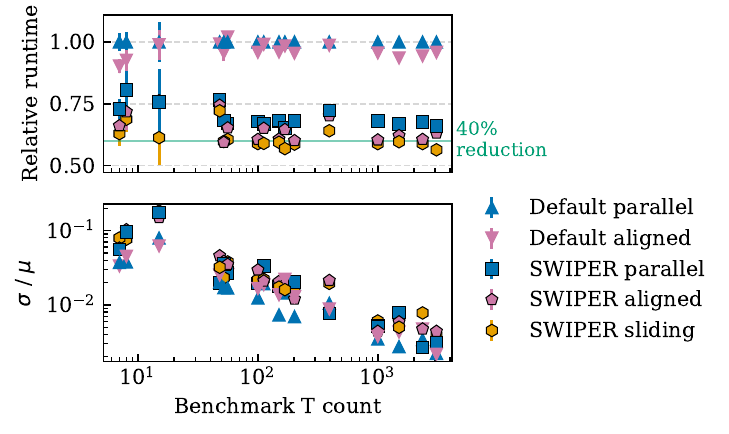}
    \caption{\emph{Top:} Performance improvement of SWIPER appears to be consistent across benchmarks size. \emph{Bottom:} Uncertainty in runtime,
    %(calculated as the coefficient of variation) versus the number of T gates for a suite of benchmark programs with different window strategies
    showing that relative uncertainty decreases for larger programs.}
    \label{fig:stdev-scaling}
\end{figure}

It's important to note that there is uncertainty in the runtime of the benchmark programs we simulate. Randomness is introduced in two ways in our simulations: \circlenumber{1} conditional S gates, which are applied $50\%$ of the time after a T gate teleportation, and \circlenumber{2} mispredictions in SWIPER. To capture this uncertainty, we run 10 trials of each benchmark with fewer than 3,000 T gates. For these benchmarks, the runtimes in Figure~\ref{fig:benchmarks-combined} have error bars showing the standard deviation of results. In Figure~\ref{fig:stdev-scaling}, we show that this standard deviation (relative to mean) decreases as T count increases while the relative improvement over the baseline remains constant. We therefore claim that, in the limit of the large T counts we expect in future fault-tolerant algorithms, SWIPER will show performance improvements similar to those in the larger benchmarks in our study. We also note that the amount of uncertainty does not appear to depend strongly on the window strategy being used, indicating that most of the variation stems from \circlenumber{1} (the conditional S gate) rather than \circlenumber{2} (mispredictions in SWIPER). This can be explained by the fact that conditional S gates occurs with $50\%$ probability whereas mispredictions occur only with $10\%$ probability. 

Looking forward, because SWIPER's main benefit is reducing the reaction time for T gates, which are typically the main cost of a program~\cite{litinski2019game, beverland2022assessing}, the magnitude of runtime improvement is broadly independent of program size, so we expect we would see similar improvements for full application-level benchmarks.

% Furthermore, the standard deviations in program duration diminish as program size increases. Figure \ref{fig:stdev-scaling} shows 

%\textcolor{red}{discuss fig. 16 and title appropriately}
\subsubsection{Limiting classical processors}\label{sec:classical_limit}

While in the majority of this work we assume the number of classical decoders is not limited, in practice we must instantiate a finite number of decoders to implement SWIPER on real hardware. As speculation failures occur probabilistically, we cannot know the exact optimal number of classical decoding processors to provision. We suggest a simple heuristic to provision an appropriate number of processors: we simulate the program of interest in the perfect-speculation, unlimited-processor case and track $P_{\text{max}}$ (maximum number of parallel processes) and $P_{\text{mean}}$ (mean number of parallel processes) over all rounds of the program. We then allocate $P_{\text{max}} + \epsilon_{\text{spec}}P_{\text{mean}}$ processors to SWIPER, where $\epsilon_{\text{spec}}$ is the probability of a misprediction ($10\%$ in our evaluation). The intuition for this heuristic is that $P_{\text{max}}$ is the number of processors required by the program and $\epsilon_{\text{spec}}P_{\text{mean}}$ is an expected number of processors needed to handle mispredicted windows that are being redecoded.

\begin{figure}
    \centering
    \includegraphics[width=\linewidth]{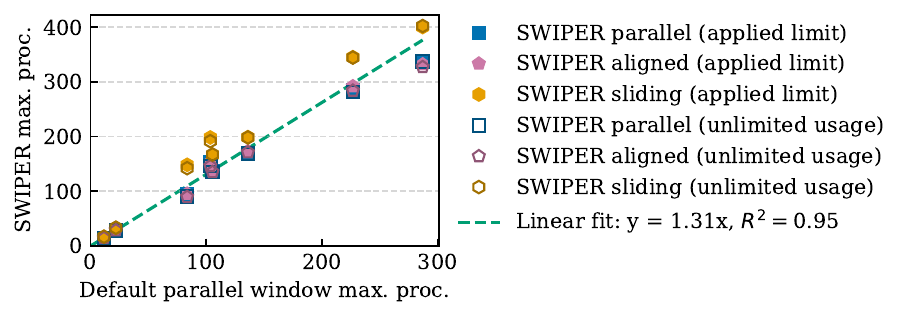}
    \caption{Comparing classical compute cost between baseline method and SWIPER.}
    \label{fig:processor-counts}
\end{figure}

%The histogram in the top right of Figure~\ref{fig:benchmarks-combined} shows the relative difference in runtime for all SWIPER trials when processor count is limited in this way. The limit appears to have minimal effect; runtimes are generally within the standard deviation of the unlimited runtimes. 
We do not observe any significant variation in simulated program runtime with this limit imposed.
Figure \ref{fig:processor-counts} compares the unlimited-processor usage to this auto-set processor limit, showing that the heuristic is able to accurately estimate the true number of required processors without bottlenecking the decoder, setting the processor limit very near to the actual maximum required. A linear fit to this data also reveals that SWIPER uses approximately 31\% more simultaneous decoding processors than the default method. We argue that this extra cost is worth the significant improvement in quantum program runtime, as classical hardware is inexpensive compared to a large-scale quantum computer.

% The uncertaintySince SWIPER has a Figure \ref{fig:benchmarks-relative} also shows the impact of limiting classical parallelism

\section{Related Work}
While QEC decoding has broadly been an active area of research~\cite{ueno2021qecool, das2022afs, byun2022xqsim, maurya2024managing}, related work on decoder performance has largely focused on achieving a decoding latency within \SI{1}{\us} as a requirement for real-time decoding, motivated by the fact that superconducting systems generate a round of syndrome measurements every $\sim$ \SI{1}{\us}~\cite{acharya2024quantum}. 
LILLIPUT~\cite{das2022lilliput} proposes a look-up table that can decode up to $d=5$ within \SI{1}{\us}. Astrea~\cite{vittal2023astrea} extends this to $d=9$ in \SI{1}{\us} via brute-force searching low-Hamming-weight bitstrings. 
Clique~\cite{ravi2023better} is a lightweight pre-decoder that specifically tackles isolated errors and bears some resemblance to the 1-Step Predictor proposed in this work.
More recently, Promatch~\cite{alavisamani2024promatch} proposes an adaptive predecoding step to lift this limit to $d=13$ within \SI{1}{\us} in regimes of low physical error rate ($0.01\%$). Barber et al.~\cite{barber2023real} design a Collision Clustering decoder and an FPGA implementation that can reach $d=23$ in under \SI{1}{\us}. Helios~\cite{liyanage2023scalable} demonstrates an impressive implementation of the Union-Find decoder~\cite{delfosse2021almost} on an FPGA and achieves latencies of \SI{11.5}{ns} for one round of a $d=21$ surface code under a simpler, phenomenological error model. 
While all of these works improve decoding latencies, the advent of parallel window decoders presents a scalable ``outer-level'' solution to real-time decoding that removes \SI{1}{\us} as a hard requirement for decoding latency. By relaxing the constraints on the inner decoder, this also enables decoders which historically have had prohibitive latencies, such as higher accuracy decoders~\cite{bausch2023learning, shutty2024efficient} and decoders for qLDPC codes~\cite{panteleev2021degenerate}.

Prior work specifically addressing parallel window decoders is still limited. The original proposals for parallel window decoding~\cite{skoric2023parallel, tan2023scalable}, work analyzing their performance for spatial windows during lattice surgery~\cite{lin2024spatially}, and work extending this to transversal gate computation for high-connectivity systems~\cite{zhang2024integrating} all assume windows with dependencies wait until their dependencies are completely decoded. SWIPER, however, proposes key differences. Our novel introduction of a speculation step allows us to reduce the reaction time of T gates and in turn fault-tolerant program runtimes by 40\%. 
%Speculation is performed by SWIPER's predictor, which is able to scalably predict data dependencies between surface code windows with $90\%$ accuracy at physical error rates of $0.1\%$.
We are also the first work to simulate general lattice surgery programs in the context of window decoding. 
%and we discover the importance of T gate alignment in reducing reaction time and propose a modified window schedule that enforces good alignment of T gates. 
\section{Conclusion and Future Work}

In this work we proposed SWIPER, a parallel window decoder that introduces a light-weight speculation step to resolve data dependencies between adajcent surface code decoding windows. There are a number of interesting directions for future work to explore. While in this work we design an independent predictor, exploring whether iterative decoding algorithms~\cite{higgott2023sparse, wu2023fusion} could admit a high-quality prediction from an intermediate state could further reduce classical resources. Additionally, SWIPER focuses on surface codes, but parallel window decoding is also compatible with a broader set of codes, including qLDPC codes~\cite{skoric2023parallel}. Designing a predictor in this setting could further extend the benefits of SWIPER.

%Through a state-of-the-art compilation pipeline and a novel round-level window decoder simulator, we find SWIPER reduces T gate reaction times and resulting program runtimes by $\sim 40\%$ on average compared to prior parallel window decoders. \jason{interesting things to discuss: (1) using first few steps of growth-based decoder as speculation (2) allowing for slower decoders at fixed reaction time}
\section*{Author contributions}

J.V. conceived of the idea, designed the 3-step predictor, designed the aligned scheduling strategy, and wrote the compilation pipeline. J.D.C. developed SWIPER-SIM and ran benchmark simulations. S.J. performed FPGA evaluations of the predictor. G.S.R., Y.L., and F.T.C. advised the project. All authors revised the manuscript.
\section*{Acknowledgements}
We thank Pranav Gokhale, Kevin Gui, and Tina Oberoi for feedback on an earlier version of this work.

This work is funded in part by EPiQC, an NSF Expedition in Computing, under award CCF-1730449; in part by STAQ under award NSF Phy-1818914/232580; in part by NSF award 2340516; in part by the US Department of Energy Office of Advanced Scientific Computing Research, Accelerated Research for Quantum Computing Program; and in part by the NSF Quantum Leap Challenge Institute for Hybrid Quantum Architectures and Networks (NSF Award 2016136), in part based upon work supported by the U.S. Department of Energy, Office of Science, National Quantum Information Science Research Centers, and in part by the Army Research Office under Grant Number W911NF-23-1-0077. This work was completed in part with resources provided by the University of Chicago’s Research Computing Center. The views and conclusions contained in this document are those of the authors and should not be interpreted as representing the official policies, either expressed or implied, of the U.S. Government. The U.S. Government is authorized to reproduce and distribute reprints for Government purposes notwithstanding any copyright notation herein. FTC is the Chief Scientist for Quantum Software at Infleqtion and an advisor to Quantum Circuits, Inc.
\section*{Data Availability}

This manuscript is currently under peer review. We will fully open-source the code and supporting data when it is published.

\bibliographystyle{ACM-Reference-Format}
\bibliography{references}

\end{document}